\documentclass[11pt]{article}
\RequirePackage{a4}         %  
\RequirePackage{times}      %
\usepackage{makeidx,showidx}
\usepackage{fancyheadings}
\usepackage{amsmath}  
\usepackage{latexsym}
\usepackage{amssymb}
\usepackage{pifont}
\usepackage{fancybox}
\usepackage{epsfig}
\usepackage{rotating}
\usepackage[english]{babel}
\Pifont{pzd}

\newcommand{\bm}[1]{\mbox{\boldmath$#1$}}

\makeatletter    % @ is a letter (instead of a special symbol )
\long\def\@makecaption#1#2{%
  \vskip\abovecaptionskip
  \sbox\@tempboxa{{\sf #1}: #2}%        % modified
  \ifdim \wd\@tempboxa >\hsize
     {\sf #1}: #2\par                   % modified
  \else
    \hbox to\hsize{\hfil\box\@tempboxa\hfil}%
  \fi
   \vskip\belowcaptionskip}
\makeatother     % @ is other
\renewcommand{\sectionmark}[1]%
                  {\markright{\thesection\ #1}}
\newdimen\larghezza
\newdimen\marginepari
\newdimen\separazione   % serve per ridefinire la separazione fra 
\def\be{\begin{equation}}
\def\ee{\end{equation}}
\def\bea{\begin{eqnarray}}
\def\eea{\end{eqnarray}}

\def\beq{\begin{equation}}
\def\eeq{\end{equation}}
\def\mvec#1{{\bm{#1}}}

\title{Bayesian model comparison applied to the 
Explorer-Nautilus 2001 coincidence data}
\author{ {\bf P. Astone,$^1$ G. D'Agostini,$^1$ S. D'Antonio$^2$}\\
1) INFN and 
University of Rome ``La Sapienza'', Rome, Italy\\
2) INFN and 
University of Rome ``Tor Vergata'', Rome, Italy\\
(Presented by P. Astone)}
\begin{document}
\setcounter{tocdepth}{1}
%%%%%%%%%%%%%%%%%%%%%%%%%%%%%\frontmatter
\pagestyle{fancy}
\date{}
\maketitle

\abstract{
\noindent
Bayesian reasoning is applied to the data by the ROG
Collaboration, in which gravitational wave (g.w.) signals are searched for 
in a coincidence experiment between Explorer and Nautilus. 
The use of Bayesian reasoning allows,
under well defined hypotheses, even tiny pieces of evidence in favor 
of each model to be extracted from the data. The combination 
of the data of several experiments can therefore be performed
in an optimal and efficient way.
Some models for Galactic sources are considered and, within 
each model, the experimental result is summarized with the 
likelihood rescaled to the insensitivity limit value (``${\cal R}$ function'').
The model comparison result is given in 
in terms of Bayes factors, which quantify how the ratio of beliefs about 
two alternative models are modified by the experimental observation.

\section{Introduction}
A recent analysis of data from the resonant g.w. detectors Explorer and
Nautilus~\cite{CQG02} has shown some hints of a possible signal over
the background expected from random coincidences. 
The indication appears only when the data are analyzed as a
function of the sidereal time. 
 Reference \cite{CQG02} does not contain
statements concerning the probability that some of the observed
coincidences could be due to g.w.'s rather than background. 
Only bottom plots of Fig.~5 and Fig.~7 of that paper gives p-values 
(the meaning of `p-value`, to which physicists are not accustomed, will
be clarified later) for each bin in sidereal
time, given the average observed background at that bin.
But p-values are not probabilities that the `only background' 
hypothesis is true, though they are often erroneously taken as such, leading to
unpleasant consequences in the interpretation of the 
data~\cite{giulioWS}.
Indeed, in this case too, Fig.~5 and Fig.~7 of Ref.~\cite{CQG02} might have 
produced in some reader sentiments different from those of the
members of the ROG Collaboration, 
who do not believe with high probability to have observed g.w.'s. 
However, the fact remains that the data are somewhat intriguing,
and it is therefore important to quantify how much we can reasonably
believe the hypothesis that they might contain some g.w. events.
The aim of this paper is to show how to make a quantitative assessment
of how much the experimental data prefer the different models in hand.

The choice of the Bayesian approach is quite natural to
tackle these kind of problems, in which we are finally interested
in the comparison of the probabilities that different models
could explain the observed data. In fact, the concept of probability
of hypotheses, probability of `true values', probability of causes, etc., 
are only meaningful in this approach. The alternative 
(`frequentistic') approach forbids to speak  about 
probability of hypotheses. Frequentistic `hypothesis test' results are given
in terms of `statistical significance', a concepts which notoriously
confuses most practitioners, since it is commonly (incorrectly!) interpreted
as it would be the probability of the `null hypothesis'~\cite{giulioWS}.
Moreover, this approach provides 
only `accepted/rejected' conclusions, and thus it is
not suited to extract evidence from noisy data 
and to combine it with other evidence provided by other data.

In the next section we present shortly the experimental data, referring
to  Ref.~\cite{CQG02} and references therein for details. 
%In Sec.~\ref{sec:p-values}
Then we review how the problem is approached 
in conventional statistics, explaining the reasons why we think
that is unsatisfactory. 
%In Sec.~\ref{sec:Bayes} 
Finally, we illustrate the Bayesian 
alternative for parametric inference and model comparison, and apply it
to the ROG data. 
%A discussion of the results and conclusions follow.

\section{Experimental data}
\begin{figure}
\begin{center}
\epsfig{file=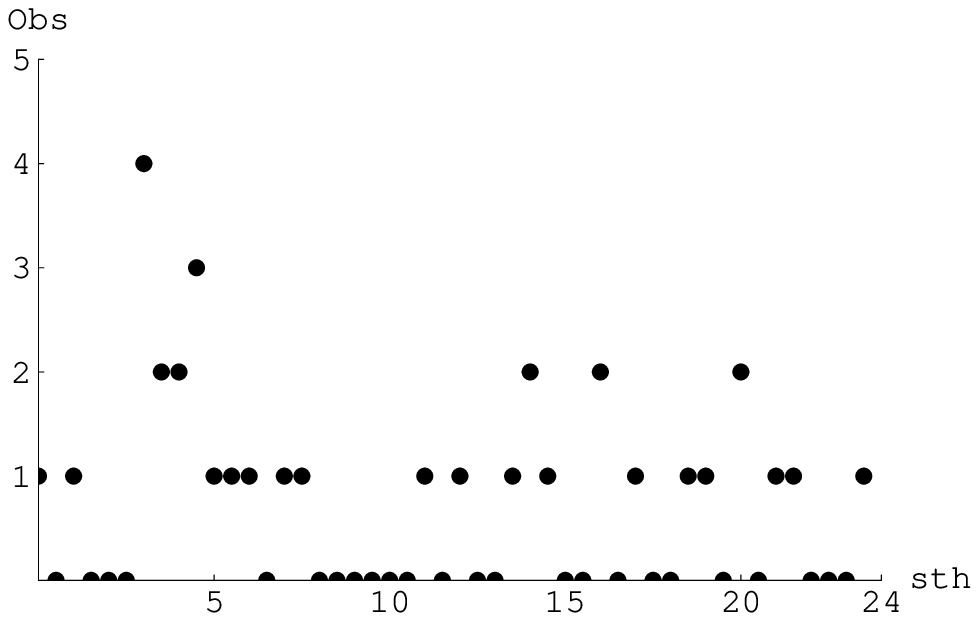,clip=,width=8cm}
\epsfig{file=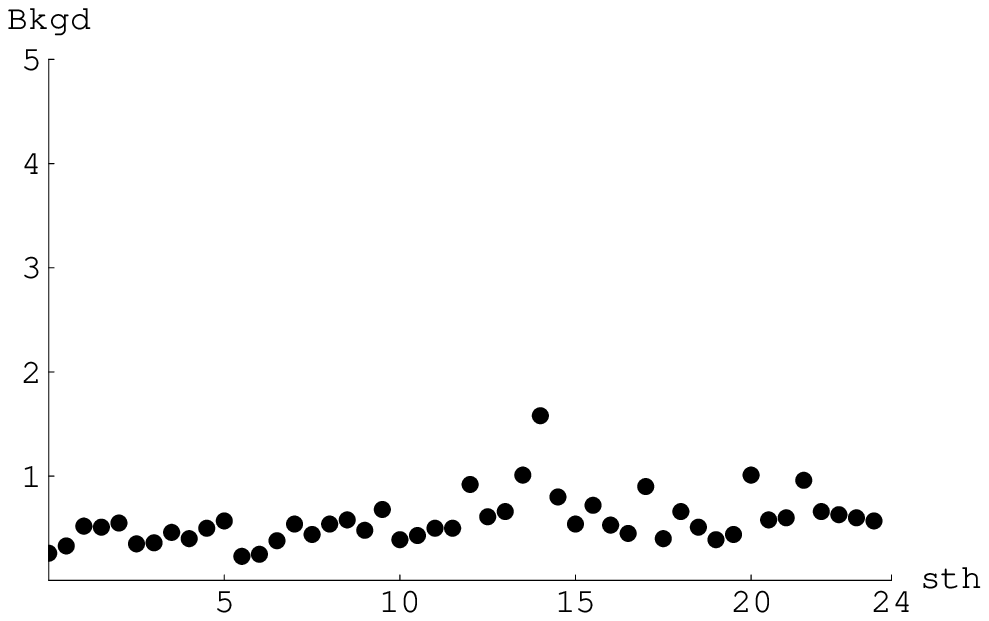,clip=,width=8cm}
\caption{Explorer-Nautilus coincidence events (upper plot) and 
background estimates (lower plot)
as a function of the sidereal time in 1/2 hour bins.}
\label{fig:figdati}
\end{center}
\end{figure}
This analysis has been performed on a data set of Explorer-Nautilus 
coincidences with an energy filter veto
(i.e. requiring agreement between the event energies of the two antennae)
and with a fixed time window of $\pm 0.5$ s.
Data we are referring to are 
those obtained using runs longer than 12 hours. 
The data are grouped in 
half hour bins of
sidereal time, as shown 
% We use here a bin width of  half hour, 
%thus our plots start from 0, which corresponds
%to the range (0:00--0:30), and ends at 23.5, which 
%corresponds to the range (23:30--24:00).
%The total number of bins is 48. The data are shown 
in  Fig.~\ref{fig:figdati}.  
The upper plot of the figure reports
the number of observed coincidences ($n_c$), while the lower plot
gives the average number of the background events 
estimated by off-time techniques.
It is worth remarking that the method we are going to use does not 
depend critically
on the width of the bins, provided that the width
is small enough to assure a good resolution of the antenna pattern. 
(To state it clearly, contrary to other methods in which some binning
is required and the resulting significance 
depends dramatically on the choice of the binning, 
in our method we could have, virtually,
bins of arbitrary small width. Rebinning
does not spoils the quality of the information, as long as 
the binning is finer than the structures exhibited by the 
antenna pattern
%of the coincidence experiment 
and there are no clustering of events
within a bin. The latter possibility is excluded by inspecting the 
arrival time of the individual events, as shown in Ref.~\cite{CQG02} 
for the events around 4:00.)

As far as the background is concerned, we recall that the random
coincidence 
background is well described by a Poisson 
distribution~\cite{CQG02}, 
and that the sidereal hour fluctuations of the averages
is compatible with the grand average over the 24 hours
of $0.57\pm 0.03$ events/hour. For these reasons, we believe that
the the value of $\lambda_B = 0.57$ is the most reasonable value
to use as parameter of the Poisson distribution which models
the background fluctuation in the 0.5-hour bins. 

\section{P-value analysis of the `statistical significance' of the data}
\label{sec:p-values}
P-value is the term preferred in modern statistics to
describe what physicists call, in simple words, ``probability of the tail(s),'' or
``probability to observe the events actually observed, 
or rarer ones, {\it given} a null hypothesis'' 
(note `given': the probability
of whatever {\it has been} observed, without the specification of a 
particular condition, is always unity). 
In the frequentistic approach,
the null hypothesis is rejected with a significance level $\alpha$ if 
the p-value gets below $\alpha$, where $\alpha$ is typically chosen to be $5\%$
or $1\%$. Besides the recognized misinterpretation of the p-value result
(see e.g. \cite{giulioWS}), there are often disputes about how this
reasoning should be applied, because it is easy to show that 
there is much arbitrariness in the kind of test to be performed 
(it is well known that practitioner often seeks for the test that tells 
what they like, moving for $\chi^2$-test, to run-test and to 
other tests with fancy names, if the previously tried tests
were ``not sensitive to the effect'') and 
in the data to include in the test,
as it is sketched in the following subsections.

\subsection{P-value based on the overall number of events}
The expected number of events due to the null hypothesis 
$H_0 = \mbox{``only background''}$ is 27.4 ($= 0.57\times 48$).
Having observed 34 events we get:
\begin{equation}
\left.\mbox{p-value}\right|_{\mbox{integral}} 
 = P(n_c \ge 34\,|\,{\cal P}_{\lambda_B=27.4}) = 12\%\,,
\label{eq:p-val_global}
\end{equation}
a value that it is not considered `significative'. However, the obvious
criticism to this procedure is that we have only used the integrated number
of coincidences,
losing completely the detailed information provided by the time distribution. 
The problem can be better understood in the limiting case of 1000 bins, 
an expected background of 1 event/bin, and an experimental result in which 999 bins
have contents which `nicely' (Poisson) fluctuate around 1, and a single bin 
exhibiting a spike of 31 counts. The p-value would be of 16\%, accepting
the hypothesis that the data are explained well by background alone.   

\subsection{P-value based on the bin presenting the highest excess}
The na\"\i ve solution to this paradox is to calculate a p-value
using only the bin presenting the highest fluctuation. This approach would give 
a very small number ($0.5\times 10^{-36}$) in our 1000 bin example, and would
remain below the  $1\%$ threshold even if the spike has only 5 events over
a background of 1. 
Applying this reasoning to the ROG data we get 
\begin{equation}
\left.\mbox{p-value}\right|_{\mbox{max}} = 
P(n_c \ge 4\,|\,{\cal P}_{\lambda_B=0.57}) =2.8 \times 10^{-3} \,,
\label{eq:p-val_peak_m}
\end{equation}
a p-value which may be considered `significative'. (Note that, if we used the observed
background of Fig.~\ref{fig:figdati}, that we do not believe it is the correct number
to use, the p-value would be $5.2 \times 10^{-4}$ .)

\subsection{P-value based on the argument that the highest excess could have shown up
everywhere in the time distribution}\label{sec:p-values_scan}
Again, the previous procedure can be easily criticized, 
because ``the bin to which the test has been applied has been chosen
after having observed the data, while a peak would have been arisen, 
a priori, everywhere in the plot.'' The standard procedure to overcome
this criticism is to calculate the probability that a peak of
that or higher value would have shown up everywhere in the data, i.e.
\begin{equation}
\left.\mbox{p-value}\right|_{\mbox{scan}} = 
1-\prod_{i=1}^{n_{bin}} F(3\,|\,{\cal P}_{\lambda_B})\,,
\label{eq:p-val_peak_scan}
\end{equation}
where $n_{bin}$ is the number of bins and $F(\cdot)$ stands for the cumulative
distribution (the product in Eq.~(\ref{eq:p-val_peak_scan})
is based on the assumption of independence of the bins).
In our case we 
get 13\% or 23\% depending whether a constant or varying background 
is assumed, i.e. p-values above any over-optimistic choice of the p-value threshold.

It is interesting to note that the $13\%$ p-value can be reobtained 
approximately as 
$\left.\mbox{p-value}\right|_{\mbox{scan}} 
\approx n_{bin}\times \left.\mbox{p-value}\right|_{\mbox{max}}$, showing 
that even a very pronounced excess can be considered not significant if 
a large number of observational 
bins are involved in the experiment (and practitioners restrict arbitrary
the region to which the test is applied, if they want the test to state
what they would like\ldots). The dependence of the result of the method 
on observations far from the 
region where there could be a good physical reason to have a signal is
annoying (and for this reason, practitioners who choose 
a suitable region around the peak do, intuitively, something correct\ldots). 
On the other hand, the reasoning does not take into account that other
bins could be interested by the signal. 

We shall see in Sec.~\ref{sec:model_comparison} how to 
use properly the prior knowledge that a (physically motivated) 
signal could have appeared everywhere in the histogram. 

\subsection{Why not to use p-values}
To conclude this section, let us summarize the reasons for
not to use procedures based on p-values.
\begin{itemize}
\item 
The interpretation of p-values is misleading, because they do
not provide probabilities of hypotheses, 
though they sound and are commonly interpreted as such. 
\item
Methods based on p-values pretend to provide answers only based
on the statistical properties of the null hypothesis, without
taking into account if other hypotheses are conceivable, and how the
alternative hypotheses describe the data. For example, 
these methods do not take into account the fact that 
a supposed signal appears at a given place rather than elsewhere,
which bins could be affected by a physical model
and how reasonable a model is.
\item 
These methods provide only binary answers, accepted/rejected.
As a consequence they are not efficient enough to analyze 
rare phenomena, which can only be discovered by
a proper combination of (even very) small pieces of evidence.
\end{itemize}

\section{The Bayesian way out: how to use of experimental data
to update the credibility of hypotheses}\label{sec:Bayes}
We think that the solution to the above problems consists
in changing radically our attitude, instead of seeking for
new `prescriptions' which might cure a trouble but generate others.
The so called Bayesian approach, based on the natural idea
of probability as `degree of belief' and on the rules of logic,
seems to us to be the proper way to deal with our problem.
A key role in this approach is played by Bayes' theorem,
which, apart from normalization constant, can be stated as
\begin{equation}
P(H_i\,|\,\mbox{\it Data}, I_0) \propto P(\mbox{\it Data}\,|\,H_i,I_0) \cdot
      P(H_i\,|\,I_0)\,,
\label{eq:BayesTh}
\end{equation}
where $H_i$ stand the hypotheses that could
produce the {\it Data}  with {\it likelihood} $P(\mbox{\it Data}\,|$ $H_i,I_0)$.
$P(H_i\,|\,\mbox{\it Data}, I_0)$ and
$P(H_i\,|\,I_0)$ are, respectively, 
the  {\it posterior}  and {\it prior} probabilities, i.e. with or 
without taking into account 
the information provided by the  {\it Data}. $I_0$ stands 
for the general status of information, 
which is usually considered implicit
and  will then be omitted in the following formulae. 

The presence of priors, considered a weak point by opposer's
of the Bayesian theory, is one of the points of force of the theory.
First, because priors are necessary to make the `probability
inversion' of Eq.~(\ref{eq:BayesTh}).
Second, because in this approach all relevant conditions must be clearly stated,
instead of being hidden in the method or left to the arbitrariness 
of the practitioner. Third, because prior knowledge can be properly
incorporated in the analysis to integrate missing or deteriorated
experimental information (and whatever it is done should be 
stated explicitly!). Finally, because the clear separation 
of prior and likelihood in Eq.~(\ref{eq:BayesTh}) allows to 
publish the results in a way independent from $P(H_i\,|\,I_0)$,
if the priors might differ largely within the members of the
scientific community. In particular, the Bayes factor, 
defined as 
\begin{equation}
BF_{ij} = \frac{ P(\mbox{\it Data}\,|\,H_i) }{ P(\mbox{\it Data}\,|\,H_j) }\,,
\label{eq:BF}
\end{equation}
is the factor which changes the `bet odds' (i.e. probability ratios)
in the light of the new data. In fact, dividing member to member
 Eq.~(\ref{eq:BayesTh}) written for hypotheses $H_i$ and $H_j$, we get
\begin{equation}
\mbox{posterior odds}_{ij} = BF_{ij} \cdot \mbox{prior odds}_{ij}  \,.
\label{eq:oddsij}
\end{equation}
Since we shall speak later about models ${\cal M}_i$, the odd ratio updating is given 
by 
\be
\frac{P({\cal M}_i\, |\, \mbox{\it Data})} {P({\cal M}_j\, |\,\mbox{\it Data})} =
\underbrace{
   \frac{P(\mbox{\it Data}\, | \,{\cal M}_i)} 
     {P(\mbox{\it Data}\, | \, {\cal M}_j)}
}_{\mbox{\it Bayes factor}}\,
\cdot 
\frac{P_\circ({\cal M}_i)} {P_\circ({\cal M}_j)}
\label{eq:test1}
\ee
Some general remarks are in order. 
\begin{itemize}
\item
Conclusions depend only on the observed data and on the previous knowledge. 
In particular they do not depend on unobserved data which are rarer than the
data really observed (that is what p-values imply).
\item
At least two models have to be taken into account, and the likelihood
for each model must be specified.
\item 
There is no need to consider `all possible models', 
%%(for which we can only
%%wait the end of Humanity or of other intelligent beings\ldots)
since what matters are relative beliefs.
\item
Similarly, there is no need that the model must be declared before the data
are taken, or analyzed.
What matters is that the {\it initial} beliefs should be based on general
arguments about the plausibility of each model and on agreement with 
other experimental information, 
other than {\it Data}. 
\end{itemize}
%\noindent
An analogue of Eq.~(\ref{eq:BayesTh}) applies to the parameters of a model. 
For example, if, given a model ${\cal M}$, we are interested to the 
rate of g.w. on Earth, $r$, Bayes' theorem gives  
\begin{equation}
f(r \, |\, {\it Data}, {\cal M}) \approx f({\it Data}\, | \,r, {\cal M}) 
\times f_\circ(r,{\cal M})\,,
\label{bayes1}
\end{equation}
where $f()$ stand for probability density functions (pdf)
Also in this case, a prior independent way of reporting the result 
is possible. The difficulty
of dealing with an infinite number of Bayes factors 
(precisely $\infty^2$, given each $r_i$ and $r_j$) can be overcome
defining a function ${\cal R}$
of $r$ which gives the Bayes factor with
respect to a reference $r_\circ$. This function is particularly
useful if $r_\circ$ is chosen to be the asymptotic value
at which the experiment looses completely sensitivity. 
For g.w. search this asymptotic value is simply $r\rightarrow 0$.
In other cases it could be an infinite particle mass~\cite{GiulioPeppe}
or an infinite mass scale~\cite{zeus_ci}. In the case of  g.w. rate $r$,
extensively discussed in Ref.~\cite{GiulioPia}, we get
\begin{equation}
{\cal R}_{{\cal M}}(r) = \frac{f({\it Data}\, | \,r, {\cal M})}
                   {f({\it Data}\, | \,r=0, {\cal M})}
= \frac{{\cal L}_{\cal M}(r)}{{\cal L}_{\cal M}(r=0)}\,,
\label{eq:Rr}
\end{equation}
where ${\cal L}_{\cal M}(r)$ is the model dependent likelihood.
[Note that, indeed, in the limit of $r\rightarrow 0$ the likelihood depends 
only on the background expectation and not on the specific model. Therefore
${\cal L}_{\cal M}(r=0) \rightarrow {\cal L}_{\cal M_\circ}$, where 
${\cal M_\circ}$ stands for the model ``background alone''.]
This ${\cal R}$ function has the meaning of 
{\it relative belief updating factor}~\cite{GiulioPia},
since it tells us how we {\it must} modify our beliefs of the
different values of $r$, given the observed data. In the region
where ${\cal R}$ vanishes, the corresponding values of $r$ are excluded.
On the other hand, in the region where  ${\cal R}$ is about unity, 
the data are unable to change our beliefs, i.e. we have lost sensitivity.
The region of transition between 0 and 1 defines the 
{\it sensitivity bound}, a concept that does not have a probabilistic
meaning and, since it does not refers to terms such as `confidence',
does not cause the typical misinterpretations of the frequentistic
`confidence upper/lower limits' (for a recent example of 
results using these ideas see Ref.~\cite{ROGBeppoSax}).
Values of $r$ preferred by the data are spotted by large value of ${\cal R}$.
We shall in the sequel how a plot of the ${\cal R}$ function gives 
an immediate representation of what the data tell about a parameter 
(Figs.~\ref{fig:Rall05} and \ref{fig:Rsolare}).
Another interesting feature of this function is that, if several 
independent data sets are available, each providing some information
about model ${\cal M}$, the global information is obtained multiplying 
the various ${\cal R}$ functions:
\be
{\cal R}_{{\cal M}}(r\,;\,\mbox{\it All data}) = 
\prod_i {\cal R}_{{\cal M}}(r\,;\,\mbox{\it Data}_i)\,.
\ee

\subsection{Models for Galactic sources of gravitational waves}
Having seen that the analysis has to be based on models
for the emission of g.w.'s, let us focus on some popular 
models within Galaxy. This limitation is due to the
sensitivity of the ROG detectors. The models taken into account are:
emission only from sources concentrated in the Galactic Center
(GC); emission from sources uniformly distributed over the 
Galactic Disk (GD); emission from sources distributed as the known 
visible mass of Galaxy (GMD). In addition we have also 
included the non-Galactic model of
sources isotropically distributed around Earth (ISO).
In this context, this latter model can just  be seen  as an academic example
to give a feeling of what the analysis
method would produce in such a scenario.
\begin{figure}
\begin{center}
\begin{tabular}{cc}
\epsfig{file=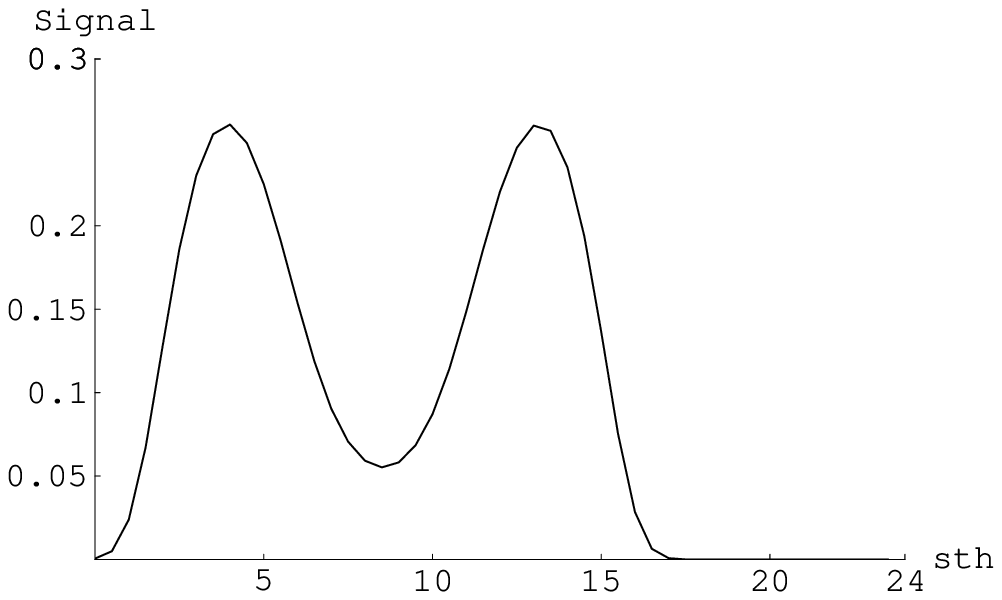,clip=,width=5.7cm} &
\epsfig{file=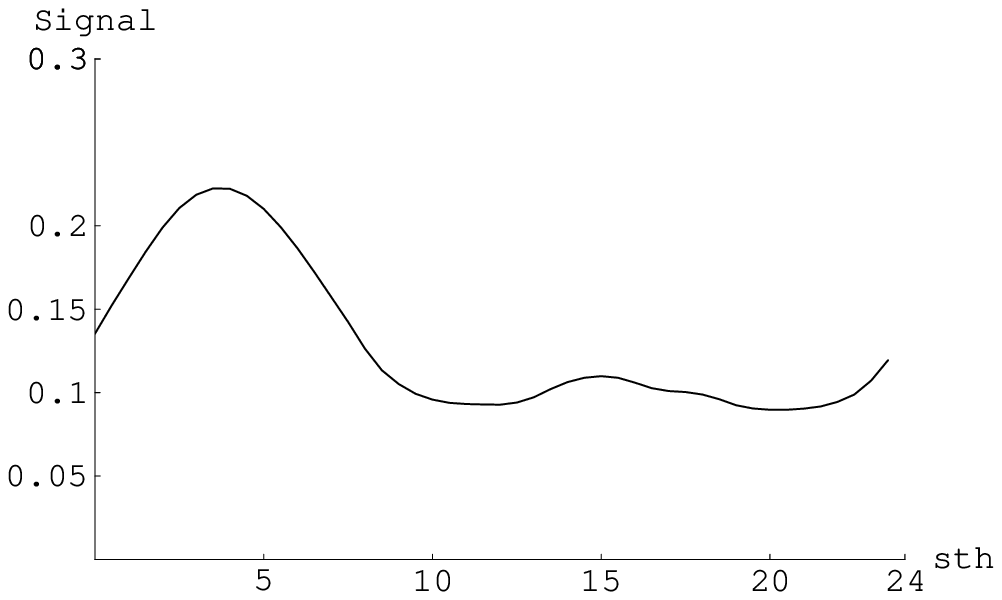,clip=,width=5.7cm} \\
Galactic Center (GC)& Galactic Disk (GD) \\
\epsfig{file=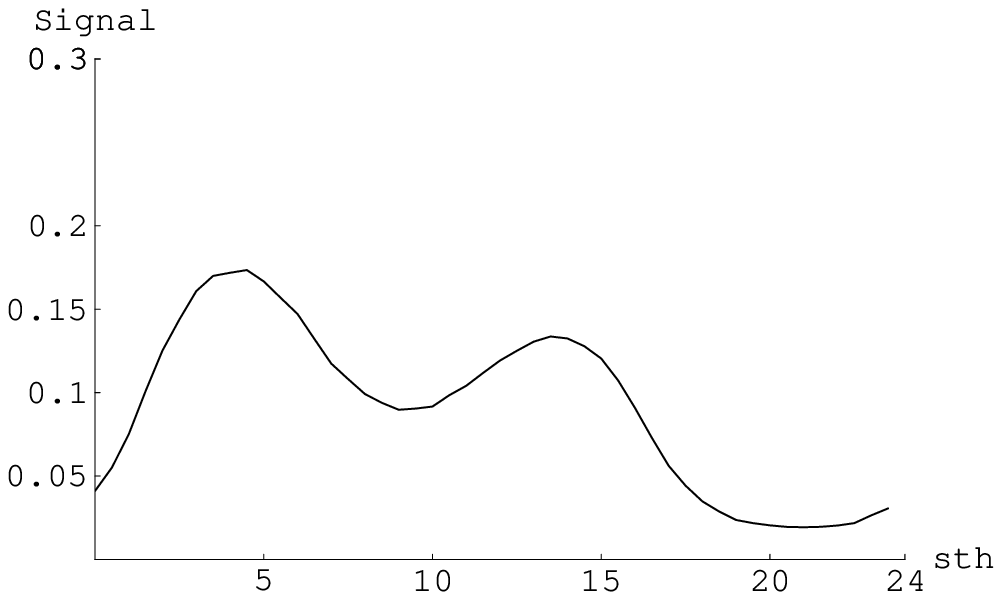,clip=,width=5.7cm} &
\epsfig{file=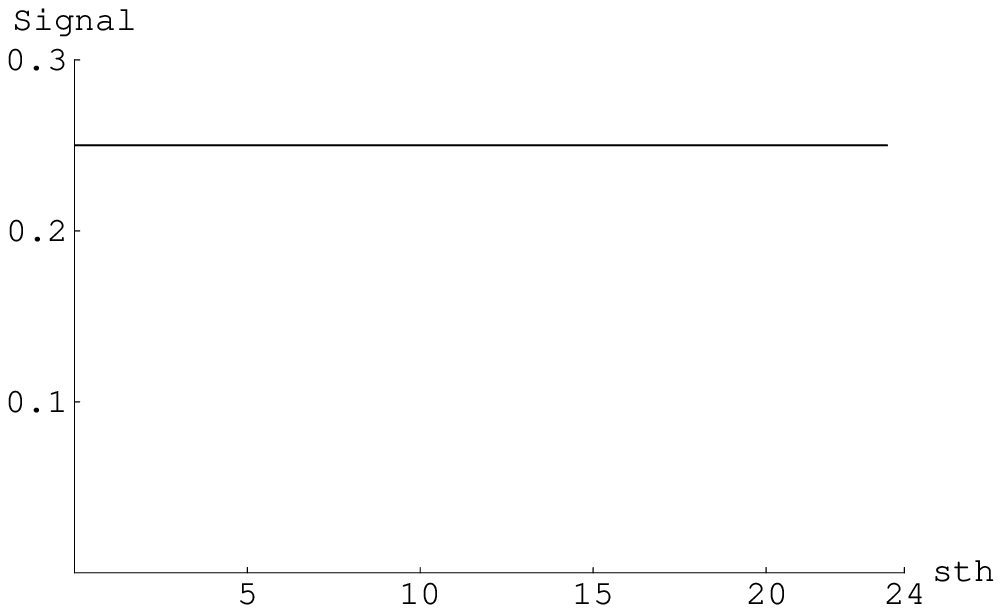,clip=,width=5.7cm}  \\ 
Galactic mass distribution  (GMD)& Isotropic (ISO)
\end{tabular}
\caption{Antenna patterns for the various models 
considered in the analysis. X-axis is the sidereal time, given in hours.
Y-axis is the antenna pattern for the coincident experiment Nautilus-Explorer,
normalized in terms of ``efficiency of detection'' (e.g. 0.25 means that, on
 the average, we need 4 signals to see 1 signal)  for signals as explained in
the text.}
\label{fig:signals}
\end{center}
\end{figure}

The pattern ${\cal P}_{{\cal M}}(i)$ of
the two detectors Explorer and Nautilus to signals due to these 
models are shown in Fig.~\ref{fig:signals}.
These patterns are all given as a function of the sidereal time. 
${\cal P}_{\cal M}(i)$ depends on the space
and energy distributions of g.w. sources described by a model, taking
also into account the variation of the 
detector efficiency with the sidereal time, due to the varying orientation 
of the antennae respect to the sources. It depends also 
on the (unknown) energies of g.w.
signals, on the noise, and on the coordinates (latitude,longitude,azimuth) 
of the detectors on Earth.
 Simply speaking, for various models of
the space distribution of sources of g.w., 
we expect different responses from the
coincidence experiment, i.e. different `antenna patterns', 
seen as a function of the sidereal time.

The calculation of ${\cal P}_{\cal M}(i)$ depends on the expected g.w. energy
and some assumptions are needed. 
The signal's amplitude $h$ is unknown, and thus we have evaluated the
patterns by integrating over a uniform distribution of $h$ values, 
ranging -- on Earth -- from $1 \times 10^{-18}$ up to  $3.0 \times 10^{-18}$.
Different reasonings may be done here, leading to
different choices for the amplitudes range and for the distributions of the signals.
We have done the simplest choice, supposing that we do not know
anything on the signals, but the fact that no signals 
have been observed at the detectors with amplitude greater than  $3.0 \times 10^{-18}$. 
The lowest limit has been
chosen considering the fact that the detector's efficiencies below $1 \times 10^{-18}$
is very small (less than $\approx 10 \%$).
Note that the considered $h$ values
are based on `standard' assumptions about the
g.w. energy release in cryogenic bars. 
The Galactic Disk (GD) model has been constructed considering g.w. sources 
uniformly distributed over the Galactic plane, which means a distribution of 
sources which is not uniform around
the Earth, given the fact that the Earth is $8.5$ kpc from the Center of
Galaxy, which is the center of the disk (whose radius is $21$ kpc).  
The GMD distribution, taking into account the mass distribution in Galaxy,
is much more interesting than the GD model. In fact we do not expect
a uniform distribution of the sources over the GD, but a distribution which is 
%highly 
concentrated near the GC \cite{paturel}. 
%%The model has been done integrating over a uniform
%%distribution of $h$ values, ranging from 
%%$0.8\times  10^{-18}$ to $3.2 \times 10^{-18}$ in the GC.
 
\subsection{Relative belief updating factor for a given model}
Having introduced the general ideas, 
let us apply them to the data of our interest.
In the $i$-th bin in sidereal time (see Fig.~\ref{fig:figdati}) 
we have $n_c(i)$ observed coincidences, with 
an expected number $\lambda_B$ due to background and 
an expected number 
\be
\lambda^{\cal M}_{S}(i) = \alpha \times {\cal P}_{\cal M}(i)
\ee
due to the detectors response to g.w. signals, which depends on the model.
The parameter  $\alpha$ is proportional to the rate $r$ of g.w. events, 
 expressed as events/day, through the following equation:

\be
\alpha \times  \sum_i{\cal P}_{\cal M}(i)=r \times T_{obs} \times \epsilon~~~~;
~~~\alpha=r \times \frac{T_{obs}} {n_{bins}}
\label{eq:alpha_eps}
\ee
where $\epsilon$ is the ``overall efficiency of detection'', $~T_{obs}$
the total observation time (which is 90 days),$~n_{bins}$ the number of bins. 
The overall efficiency of detection is calculated from the antenna pattern as
\be
\epsilon = \frac{\sum_i^{n_{bins}}  {\cal P}_{\cal M}(i)}{n_{bins}\times 1} 
\ee 
which is the ratio of the area covered by the detectors and the
area which would be covered by ideal detectors (that is, having 
${\cal P}_{\cal M}(i)$=1 at all times).
We have then the following likelihood for each model:
\be
{\cal L}_{\cal M}(\alpha) = f({\it Data}\, | \,\alpha, {\cal M}) 
= \prod_i\frac{ e^{-\lambda_{\cal M}(i)}\, \lambda_{\cal M}(i)^{n_c(i)} }
                              {n_c(i)!}\,,
\label{eq:likelihood}
\ee
with 
\be 
\lambda_{\cal M}(i) = \lambda_B + \lambda^{\cal M}_S(i)\,.
\ee 
$\lambda^{\cal M}_{S}(i)$ is the parameter of the Poisson distribution
that describes the g.w. signal given model ${\cal M}$. It can be
said, in simple words, to be the number of g.w. events
that could be present in each bin
in a coincidence experiment having 
performance and exposure time as that described in Ref.~\cite{CQG02}. 
Hence, 
\be
n_{gwc} = \alpha \times \sum_i {\cal P}_{\cal M}(i)
\ee 
gives the total number of  such g.w. events. 
This number is the sum of $\alpha$ over all the bins, 
weighted with the pattern ${\cal P}_{\cal M}(i)$.
Our interest is, via $\alpha$, 
to infer  $n_{gwc}$ and the rate $r$. However, given
the strong dependence of the inference from the priors,
typical for this kind of frontier measurements, we prefer
to report the result in terms of  ${\cal R}$ functions, as discussed
above.
\begin{figure}
\begin{center}
\epsfig{file=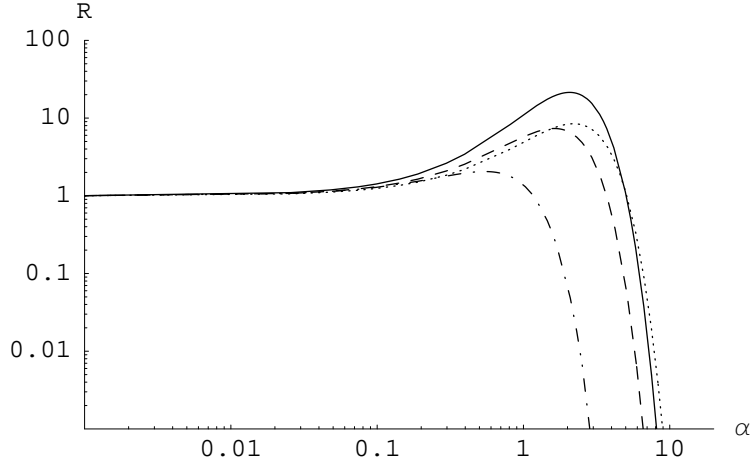,clip=,width=10cm}
\caption{${\cal R}$ function 
for the four models considered: 
Galactic Center (continuous);
Galactic Disk (dashed); GMD (dotted); isotropic (dot-dashed).}
\label{fig:Rall05}
\end{center}
\end{figure}
%Having specified the models, 
%we can evaluate the ${\cal R}$ function  ${\cal R}_{\cal M}(\alpha)$ 
%for each model. 
The resulting ${\cal R}_{\cal M}(\alpha)$'s are shown in Fig.~\ref{fig:Rall05}.
Note that the figures are in log-log scale to make it clear that many
orders of magnitude of $\alpha$ are involved.

The results are summarized in 
Tab.~\ref{tabuno05}. 
\begin{table}
\begin{center}
\caption{Summary of ${\cal R}$ function for each model, together 
with the parametric inference based on Maximum Likelihood or 
on the Bayes formula assuming a uniform prior for $\alpha$.}
\label{tabuno05}
\begin{tabular}{lcccc} 
 &&&& \\
\hline
  Model     & GC & GD & GMD & ISO\\
\hline
$\sum_i {\cal P}(i)$  &4.7 & 6.1 & 4.4 & 12\\
$\epsilon$ (\%)    &9.8 & 13 &9.2 & 25 \\
\hline 
${\cal R}_{max}$        & 21 & 7.2 & 8.2 & 1.9  \\
\hline
                       & \multicolumn{4}{c}{(events)} \\
$\alpha_{ML}         $ & 2.1 &  1.6 & 2.2 & 0.5   \\
$\sigma_{ML}(\alpha) $ & 1.0 &  0.9 & 1.2 & 0.5   \\
E$[\alpha\,|\,f_\circ(\alpha)=k]$ & 2.3 &  1.8 & 2.5 & 0.7   \\
$\sigma[\alpha\,|\,f_\circ(\alpha)=k]$       & 1.0 &  0.9 & 1.2 & 0.4   \\
\hline
                       & \multicolumn{4}{c}{(events)} \\
$n_{gwc_{ML}} $        & 10  & 10  & 10 & 7 \\
$\sigma_{ML}(n_{gwc})$ &  5  &  5  &  5 & 6 \\ 
E$[n_{gwc}\,|\,f_\circ(\alpha)=k]$           & 11  &  11  & 11  & 9     \\
$\sigma[n_{gwc}\,|\,f_\circ(\alpha)=k]$      &  5  &   6  &  5  & 5     \\ 
\hline
                       & \multicolumn{4}{c}{(events/day)} \\
$r_{ML}$               & 1.1 & 0.9 & 1.2 & 0.3 \\
$\sigma_{ML}(r)$       & 0.5 & 0.5 & 0.7 & 0.3 \\
E$[r\,|\,f_\circ(\alpha)=k]$                 & 1.2 &  1.0 & 1.3 & 0.4  \\
$\sigma[r\,|\,f_\circ(\alpha)=k]$            & 0.6 &  0.5 & 0.6 & 0.2  \\                     
\hline
\end{tabular}
\end{center}
\end{table}
$\alpha_{ML}$ is the value that maximizes ${\cal R}$. 
The symbol ML reminds that $\alpha_{ML}$ is the value that maximizes
the likelihood (likelihood and ${\cal R}$ function differ by a factor). 
Indeed, this is the result that a Maximum Likelihood (ML) analysis would produce
for $\alpha$. 
The parameter $\alpha$ is turned into g.w. rate using 
Eq.~(\ref{eq:alpha_eps}).
%the overall 
%efficiency of detection and the total effective observation time (90 days). 
%
${\cal R}_{max}$ gives how much the belief for $\alpha=\alpha_{ML}$ 
increases with respect to $\alpha=0$. The higher relative belief updating factor 
is obtained for the Galactic Center model. Within this model, 
the pdf for  $r$ 
around 1.2 events/day  
gets enhanced by a factor 21 with respect to $r=0$.

Figure \ref{fig:Rall05} shows clearly how  the initial beliefs about $\alpha$ 
(and therefore on $r$)
are updated, within each model. We want to stress that 
the final conclusion depends still on the prior beliefs.
If someone thought that $\alpha$ had to be above 10
this person had to reconsider completely his/her beliefs, 
independently from the model;
if another person believed that only values below 0.01 were
reasonable, 
the experiment would not affect at all his/her beliefs, 
independently of the model. For this reason, the ML value
$\alpha_{ML}$ could be misleading if erroneously associated, 
as it often happens, to the value around which our confidence is 
finally concentrated, independently from any prior knowledge. 
Nevertheless, and with these warnings, 
we report in Tab.~\ref{tabuno05} also the results obtained
from a ML analysis and from a 
na\"\i ve Bayesian inference that assumes a uniform prior on $\alpha$
(and therefore on $r$ and $n_{gwc}$, since they differ by factors).
$\sigma_{ML}(\alpha)$ has been evaluated from the curvature of 
the minus-log-likelihood around its minimum, i.e. 
$\sigma_{ML} ^{-2} =$ 
$\partial^2/\partial \alpha^2 \,(-\ln \,{\cal L}(\alpha))|_{\alpha_{ML}}$.
%\be
%\sigma_{ML} ^{-2}(\alpha) = 
%\left.\frac{\partial^2}
%{\partial \alpha^2}\,\left({-\ln \,{\cal L}(\alpha)}\right)
%\right|_{\alpha_{ML}}\,.
%\ee
The results of the `na\"\i ve Bayesian inference' are reported
as expected values E$[\cdot]$ and standard deviations evaluated 
from the final distribution. The condition $f_\circ(\alpha)$ has
been written explicitly in E$[\cdot]$ and 
$\sigma[\cdot]$, according to the Bayesian spirit.
 Note that, for obvious reasons, the mode of 
the posterior calculated using a uniform prior is exactly equivalent
to the ML estimate. This observation is important
to understand the slightly different results obtained with the
two methods. The posterior expected value is always larger
than the ML one, simply because of the asymmetry of ${\cal L}$. 

Perhaps $n_{gwc}$ is the most interesting quantity to understand 
the conclusions of these
{\it model dependent}
analyzes that, we like to repeat it, 
{\it do not take properly into account prior knowledge}. 
The three physical model suggest about 10 coincidences 
due to g.w.'s, with a 50\% uncertainty. Instead, 
for the unphysical model (ISO)
less events are found and with larger uncertainty. Note that, for this model,
 the mode
of the posterior (or, equivalently, the $ML$ estimate) gives a number
of candidate events that is the difference between the total number of
observed events
and that expected from the background alone. Instead, for the three
Galactic models, a number of events larger than this difference is
attributed to the signal, as a consequence of a `possibly good'
time  modulation recognized in the data
(in other words, the method `likes to think' that, given
a time distribution shape that reminds the pattern of the Galactic models, 
the background has most likely under-fluctuated within what is reasonably
allowed by its probability distribution). 

%To conclude this subsection, we can say, that, given 
%any of the three galactic models, the peak in ${\cal R}(\alpha)$
%differs slightly from model to model, in order to produce 
%a peak in the same place of ${\cal R}(n_{gwc})$. 

To summarize this subsection, 
the three Galactic 
models show good agreement in indicating for which values of
g.w. events, or event rate, we {\it must increase} our beliefs. 
But the final beliefs depend on our initial ones, 
as explained introducing the Bayesian approach.
If {\it you} think that, given your best knowledge of the models of g.w.'s sources and 
of g.w. interaction with cryogenic detectors, a g.w. rate on Earth of 
up to ${\cal O}(1)$ event/day is quite possible, the data make you
to believe that this rate is $\approx 1.0\pm 0.5$ event/day 
and that they contain $\approx 10\pm 5$ genuine g.w. coincidences.

\subsection{Model comparison taking into account the a priori possible values
of the model parameters}\label{sec:model_comparison}

While in the previous subsection we have been interested to learn about $\alpha$
or $r$ {\it within} a model (and then, since all results are conditioned by that 
model, it makes no sense from that perspective to state if the model is
right or wrong), let us see now how to {\it modify} our beliefs on each
model. 
This is a delicate question to be treated
with care. Intuitively, we can imagine that we have to
make use of the ${\cal R}$ values, in the sense that 
the higher is the value and the most `the hypothesis' 
increases its credibility. The crucial point is to understand that
`the hypothesis' is, indeed, a {\it complex} (somewhat {\it multidimensional})
hypothesis. Another important point is that, given a non null background
and the properties of the Poisson distribution, 
we are never {\it certain} that the
observations are not due to background alone (this is 
the reason why the ${\cal R}$ function does not
 vanish for $\alpha\rightarrow 0$).

The first point can be well understood  making an example based 
on Fig.~\ref{fig:Rall05} and Tab.~\ref{tabuno05}.
Comparing ${\cal R}_{ML}$ for the different models one could come
to the rash conclusion that the Galactic Center model is enhanced by 
21 with respect to the non g.w. hypothesis, or that 
the Galactic Center model  is enhanced by a factor  $21/7$ 
with respect  to the  hypothesis of signals 
from sources uniformly distributed over the Galactic Disk.
However these conclusions would be correct only in the case that each 
model would admit {\it only} that value of the parameter which
maximizes ${\cal R}$, i.e.
\begin{eqnarray}
BF_{GC(\alpha=2.1),\, GD(\alpha=1.6)} 
&=& \frac {f(\mbox{\it Data}\,|\, GC, \, \alpha=2.1)}
          {f(\mbox{\it Data}\,|\, GD, \, \alpha=1.6)} 
\label{eq:BF_M_alpha}  \\
&=&  \frac {\frac{f(\mbox{\footnotesize \it Data}\,|\, GC, \, \alpha=2.1)}
                 {f(\mbox{\footnotesize \it Data}\,|\, \mbox{\footnotesize \it BKGD alone})}
           }
           {\frac{f(\mbox{\it \footnotesize Data}\,|\, GD, \, \alpha=1.6)}
                 {f(\mbox{\it \footnotesize Data}\,|\, \mbox{\footnotesize \it BKGD alone})}
           }  
%&=&  \frac {\frac{f(\mbox{\it \footnotesize Data}\,|\, GC, \, \alpha=2.1}
%                 {f(\mbox{\it \footnotesize Data}\,|\, GC, \, \alpha=0})
%           }
%           {\frac{f(\mbox{\it \footnotesize Data}\,|\, GD, \, \alpha=1.6)}
%                 {f(\mbox{\it \footnotesize Data}\,|\, GD, \, \alpha=0)}
%           }  
= \frac{{\cal R}_{GC}(2.1)}{{\cal R}_{GD}(1.6)}
 \approx \frac{21}{7} = 3. \nonumber
\end{eqnarray}
But we are, indeed, interested in 
$BF_{GC\,, GD}$ and not in 
%the l.h.s. of 
%Eq.~(\ref{eq:BF_M_alpha}).
$BF_{GC(\alpha=2.1),\, GD(\alpha=1.6)}$.
 We must take into account 
the fact that a wide range of $\alpha$ values could be associated
to each model. 

Let us take the Bayes factor defined in Eq.~(\ref{eq:test1}). 
The probability theory teaches us promptly what to do when 
each model depends on parameters:
\be
P(\mbox{\it Data}\,|\,{\cal M}) = \int\! P(\mbox{\it Data}\,|\,{\cal M}, \mvec{\theta}) \,
 f(\mvec{\theta}) \,\mbox{d}\mvec{\theta}\,,
\ee
where $\mvec{\theta}$ stands for the set of the model parameters and 
$f(\mvec{\theta})$ for their pdf. 
Applying this formula to the Bayes factors of our interest we get
\begin{eqnarray}
BF_{{\cal M}_i, {\cal M}_j} =
\frac{P(\mbox{\it Data} \, |\, {\cal M}_i)} {P(\mbox{\it Data} \, |\, {\cal M}_j)} %& = &
%\frac
%{\int P(Data \, |\, {\cal M}_i,\,\alpha)\,f_\circ(\alpha)\,d\alpha} 
%{\int P(Data \, |\, {\cal M}_j,\,\alpha)\,f_\circ(\alpha)\,d\alpha} \\
 & = & \frac
{\int {\cal L}_{{\cal M}_i}(\alpha \, ;\, \mbox{\it Data})\,f_{\circ_{{\cal M}_i}}(\alpha)\,\mbox{d}\alpha} 
{\int  {\cal L}_{{\cal M}_j}(\alpha \, ;\, \mbox{\it  Data})\,f_{\circ_{{\cal M}_j}}(\alpha)\,\mbox{d}\alpha}
\label{eq:Rrcomplex}
\end{eqnarray}
where $f_\circ(\alpha)$ is the (model dependent) 
prior about $\alpha$. Note that the Bayes factors
with respect to ${\cal M}_0 = $''background alone'' get the simple
expression 
\be
BF_{{\cal M}, {\cal M}_0} = \int {\cal R}_{{\cal M}}(\alpha) 
\, f_{\circ_{\cal M}} (\alpha)\,\mbox{d}\alpha \,.
\label{eq:BF_R}
\ee
Equation (\ref{eq:Rrcomplex}) shows that the `goodness' of the model depends on the integrated likelihood
\be
\int{\cal L}_{{\cal M}}(\alpha \, ;\, Data)\,f_{\circ_{{\cal M}}}(\alpha)\,\mbox{d}\alpha 
\label{eq:test4}
\ee
which is sometimes called `evidence' (in the sense that ``the higher is this number,
the higher is the evidence that the data provide in favor of the model'').
It is important to note that ${\cal L}_{{\cal M}}(\alpha \, ;\, Data)$
has its maximum value around the ML point $\alpha_{ML}$, 
but Eq.~(\ref{eq:test4}) takes into account all prior 
possibilities of the parameter.
Thus, in general, it is not enough that one model fits the data
better than its alternative (think, e.g., at the minimum $\chi^2$ as a measure 
of fit goodness) to prefer finally that model.
First there are the model priors, which we have to
take into account. Second, the evidence (\ref{eq:test4}) takes into account 
the parameter space preferred by the likelihood (i.e. the values around
the ML point) with respect to the parameter space 
allowed a priori by the model. 
In the extreme case, one could have 
a model that can fit `perfectly' the experimental data after having
adjusted dozens of parameters, but this model yields a very small
`evidence' and it is therefore disregarded. This automatic filtering
against complicated models is a nice feature of the Bayesian theory and 
reminds the Ockham' Razor criterion~\cite{Berger-Jefferys}.

To better understand the role of the parameter prior 
in Eq.~(\ref{eq:test4}), let us take the example of a model 
(which we do not consider realistic and, hence, we have discarded
a priori in our analysis) that gives a signal only in one of the 1/2 hours bins,
being all bins a priori equally possible.
This model ${\cal M}_{s}$ would depend on two parameters, $\alpha$ and $t_s$,
where $t_s$ is the center of the time bin. Considering   $\alpha$ and $t_s$ independent,
the parameter prior is $f_\circ(\alpha,t_s)=f_\circ(\alpha)\cdot f_\circ(t_s)$, 
where $f_\circ(t_s)=1/48$ is a probability function for the discrete variable 
$t_s$. The `evidence' for this model would be
$$
\sum_{t_s}\int{\cal L}_{{\cal M}_s}(\alpha, t_s \, ;\, \mbox{\it Data})
\,f_\circ(\alpha)\,f_\circ(t_s)\,\mbox{d}\alpha 
=
\sum_{t_s}\frac{1}{48}\int{\cal L}_{{\cal M}_s}(\alpha, t_s \, ;\, \mbox{\it Data})
\,f_\circ(\alpha)\,\mbox{d}\alpha 
$$
If the data show a very large peak in correspondence of $t_s=t_{s_{ML}}$, 
we have that 
${\cal L}_{{\cal M}_s}(\alpha, t_{s_{ML}} \, ;
\, \mbox{\it Data}) 
\ggg {\cal L}_{{\cal M}_s}(\alpha, t_s\ne t_{s_{ML}} \, ;\, \mbox{\it Data}) 
$
and then 
$$
\sum_{t_s}\int{\cal L}_{{\cal M}_s}(\alpha, t_s \, ;\, Data)
\,f_\circ(\alpha)\,f_\circ(t_s)\,\mbox{d}\alpha 
\approx
\frac{1}{48}\int{\cal L}_{{\cal M}_s}(\alpha, t_{s_{ML}} \, ;\, Data)
\,f_\circ(\alpha)\,\mbox{d}\alpha 
$$
This model is automatically suppressed by a factor $\approx 48$
with respect to other models that do not have the time position
as free parameter. Note that this suppression 
goes in the same direction
of the reasoning described in Sec.~\ref{sec:p-values_scan}.
But the Bayesian approach tells us when and how this suppression
has to be applied. Certainly not in the Galactic models we are considering.

As we have seen, while the Bayes factors for simple hypotheses
(`simple' in the sense that they have no internal parameters) 
provide a prior-free information of how to modify the beliefs,
in the case of models with free parameters Bayes factors 
remain independent from the beliefs about the models, but do depend
on the priors about the model parameters. In our case they depend 
on the priors about $\alpha$, which might be different for different 
models. If we were comparing different models, 
each with its $f_\circ(\alpha)$ about which there is full agreement
in the scientific community, all further calculations would be 
straightforward. However, we do not think to be in such a nice
text-book situation, dealing with open problems in frontier physics
(for example, note that $\alpha$, and then $r$ and $n_{gwc}$, depend on 
the g.w. cross section on cryogenic bars, and we do not believe 
that the understanding of the underlying mechanisms is completely settled).   
In principle every physicist which have formed his/her ideas
about some model and its parameters should insert his/her functions
in the formulae and see from the result how he/she should change 
his/her opinion about the different models. 
Virtually {\it our task ends here},
having given the ${\cal R}$ functions, which can be seen as the 
best summary of an experimental fact, and having indicated how to proceed
(for recent examples of applications of this method in astrophysics and cosmology
see Refs.~\cite{LoredoLamb,John-Narlikar,Hobson}). 
Indeed, we proceed, 
showing how beliefs can change given some possible scenarios for 
$f_\circ(\alpha)$. 

The first scenario is that in which 
the possible value of $\alpha$ are considered so small that 
$f_\circ(\alpha)$ is equal to zero for $\alpha > 0.01$.
The result is simple: the data are irrelevant and beliefs
on the different models are not updated by the data. 

Other scenarios might allow
the possibility that $f_\circ(\alpha)$ 
is positive for values up to ${\cal O}(1)$ and more. We shall use  
three different pdf's for $\alpha$ as examples of prior beliefs, 
that we call `sceptical', `moderate' and 'uniform' (up to $\alpha=10$).
The `moderate' pdf corresponds to a rate
which is rapidly going to zero around the value which we have measured. 
The initial pdf is modeled with a half-Gaussian with $\sigma=1$. 
The `sceptical' pdf has a $\sigma$ ten times smaller.
The `uniform' considers equally likely
all  $\alpha$ up to the last decade
in which the ${\cal R}$ functions are sizable different from zero. 
Here are the three $f_\circ(\alpha)$:
\begin{eqnarray}
f_\circ(\alpha\,|\,\mbox{sceptical}) 
&\propto& {\cal N}(0, \,0.1) \hspace{0.5cm}(\alpha > 0) \\
f_\circ(\alpha\,|\,\mbox{moderate}) 
&\propto& {\cal N}(0, \,1) \hspace{0.75cm}(\alpha > 0) \\
f_\circ(\alpha\,|\,\mbox{uniform}) &=& k \hspace{1.72cm} (0 < \alpha < 10)\,,
\end{eqnarray}
where ${\cal N}(\mu,\sigma)$ stands for a Gaussian distribution. 
For simplicity, we use the same sets of priors for all models, though
they could, and probably should, be different for each model. 
But we think that this is sufficient for the purpose of 
this exercise, which is that of illustrating the method.  

Using these three pdf's for the parameter $\alpha$,
we can finally calculate all Bayes factors. 
We  report in Tab.~\ref{tab:due05} the Bayes factors of the models
of Fig.~\ref{fig:signals} with respect to model ${\cal M}_0 = $ ``only background'',
using Eq.~(\ref{eq:BF_R}). 
All other  Bayes factors can be calculated as ratio of these.  
\begin{table}[t]
\begin{center}
\caption{Bayes factors, 
for the four models of Fig.~\ref{fig:signals} 
 with respect to model ${\cal M}_0 = $ ``only background''
depending on three choices for  $f_\circ(\alpha)$. 
The thumbnails showing $f_\circ(\alpha)$ are log-log plots with
abscissa scales exactly as in  Fig.~\ref{fig:Rall05}.}
\label{tab:due05}
\begin{tabular}{cccc}
\hline 
 &  `sceptical'  &  `moderate'  &  `uniform' \\
$
\begin{array}{lr}  
& f_0(\alpha) \\
 & \\ 
 \mbox{Model} & 
\end{array}
$
& \epsfig{file=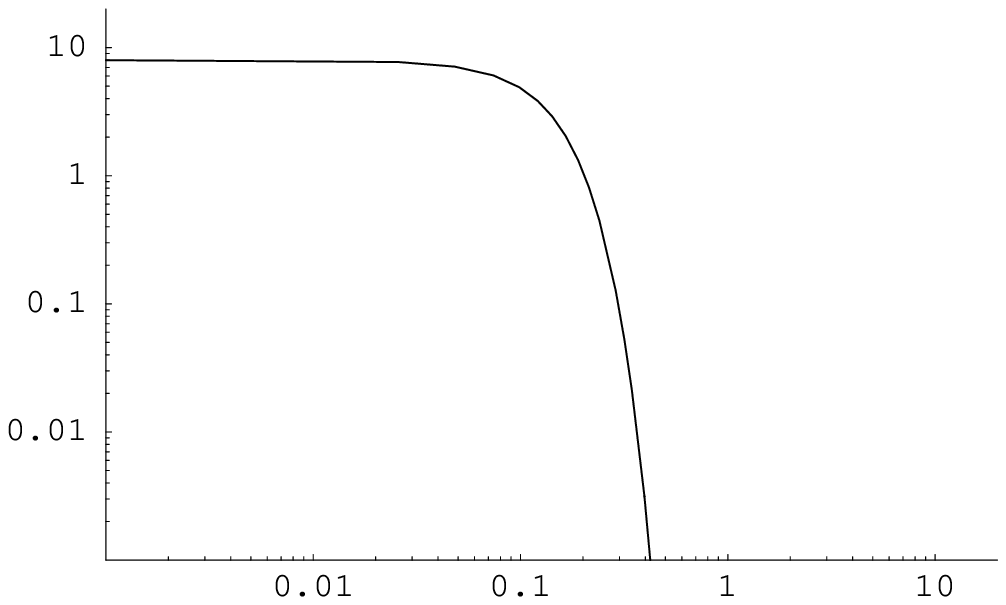,clip=,width=2.2cm,height=1.0cm} 
&\epsfig{file=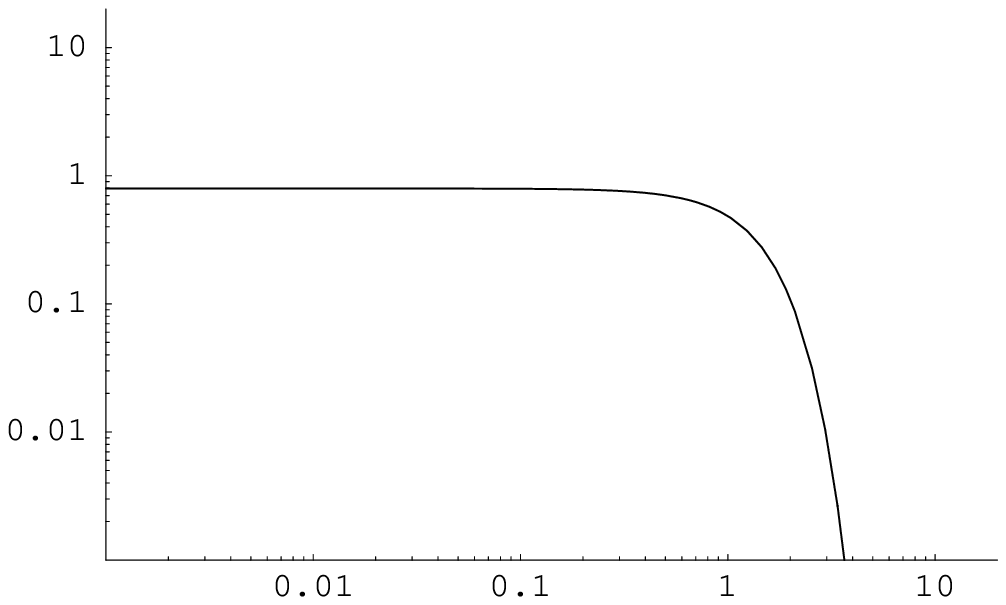,clip=,width=2.2cm,height=1.0cm}
&\epsfig{file=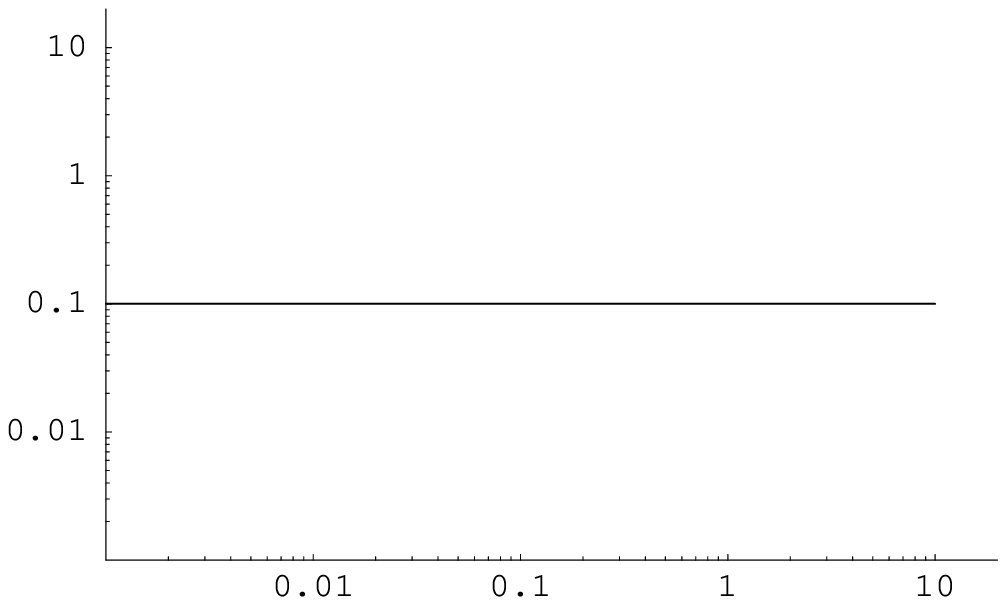,clip=,width=2.2cm,height=1.0cm}\\
\hline
%\multicolumn{1}{c}{GC} & & & \\
\epsfig{file=figlavoropub/figsignalCG.eps,clip=,width=2.5cm}  &  {\huge 1.3} &{\huge  8.4} &{\huge  5.4}   \\
\hline
%\multicolumn{1}{c}{GD} & & & \\
\epsfig{file=figlavoropub/figsignalDG.eps,clip=,width=2.5cm}   & {\huge 1.4} & {\huge 4.1} &{\huge1.7}   \\ 
\hline 
%\multicolumn{1}{c}{GMD} & & & \\
\epsfig{file=figlavoropub/figsignalM1.eps,clip=,width=2.5cm}  &  {\huge 1.2} &{\huge  3.9} &{\huge2.6}   \\
\hline
%\multicolumn{1}{c}{UNI} & & & \\
\epsfig{file=figlavoropub/figsignalUNIF.eps,clip=,width=2.5cm} &   {\huge 1.2} &{\huge  1.4} & {\huge 0.2}   \\ 
\hline
\end{tabular}
\end{center}
\end{table}
The interpretation of the numbers is straightforward, remembering
Eq.~(\ref{eq:BF}). 
If the preference was for $\alpha$ values below 0.1 (the `sceptical'),
the data produce a Bayes factor just above 1 for all models, indicating
that the experiment has slightly increased our conviction, but essentially there
is no model particularly preferred. If, instead, we think, though with low probability
that even values of $\alpha$ above 1 are possible (i.e. $r \gtrsim 0.5$ event/day), 
then Bayes factors are obtained 
which that can sizable increase our suspicion that some events could be really due 
to one of these models.\footnote{To understand the quantitative role of the 
Bayes factors, let us make some examples: 8.4 means that if someone was in serious doubt
to believe or not (i.e. $P=50\%$), in the light of the data his/her belief 
increases to 89\%. A BF of 100 would make the same person `practically sure' (99\%),
or would bring into serious doubt a sceptical person (who had an initial belief of 1\%).}
Within this `moderate' scenario there is some preference for the Galactic Center model
with a Bayes factor about 2 with respect to each other model. This result
contradict the na\"\i ve judgment based on observation of a `peak' at 
around 4:00. The response of the Bayesian comparison takes into account 
all features of the model pattern, including the width of the peaks. 
%As a numerical exercise, take a person that gives a 10\% chance that 
%either model GC or GMD are possible, and that has no preference between
%the two, while he/she does not believe at all model GD. Given each model,
%he/she `sceptical' (as modelled above) about $\alpha$. We get then the
%following initial and final odds:
%\begin{eqnarray}
%\frac{P_\circ(GC)}{P_\circ({\cal M}_\circ)} = 
%\frac{P_\circ(GMD)}{P_\circ({\cal M}_\circ)} &=& \frac{5}{90} = 0.056 \\
%\frac{P_\circ(GC)}{P_\circ(GMD)} &=&  1 \\
%\frac{P(GC\,|\,\mbox{\it Data},\,\mbox{scept.}\, f_\circ(\alpha)
%)}{P({\cal M}_\circ\,|\,\mbox{\it Data},\,
%\mbox{scept.}\, f_\circ(\alpha))} &=& 0.14 \\
%\frac{P(GC\,|\,\mbox{\it Data},\,\mbox{scept.}\, f_\circ(\alpha)
%)}{P({\cal M}_\circ\,|\,\mbox{\it Data},\mbox{scept.}\, f_\circ(\alpha)
%)} &=& 0.11 \\
%\frac{P(GC\,|\,\mbox{\it Data},\mbox{scept.}\, f_\circ(\alpha)
%)}{P(GMD\,|\,\mbox{\it Data},\mbox{scept.}\, f_\circ(\alpha)
%)} &=&  1.5\,,
%\end{eqnarray}
%from which, for example, we can calculated the probability the models, 
%under the condition than either of the two is the right one:
%\begin{eqnarray}
%P(GC\,|\,\mbox{\it Data}, \,\mbox{scept.}\, f_\circ(\alpha),\, GC\cup GMD) &=& 60\% \\
%P(GMD\,|\,\mbox{\it Data}, \,\mbox{scept.}\, f_\circ(\alpha),\,GC\cup GMD) &=& 40\%\,.
%\end{eqnarray}

We have also considered a prior which is uniform in $log(\alpha)$, between $\alpha=(0.001-10)$.
This prior accords equal probability to each decade in the parameter $\alpha$, and probably
accords many people prior intuition.
Bayes factors, for the four models of Fig.~\ref{fig:signals} 
 with respect to model ${\cal M}_0 = $ ``only background'', are:

4.0 (GC); 2.0 (GD); 2.2 (GMD); 1.0 (ISO). 

Again, within this scenario there is some preference for the Galactic Center model
with a Bayes factor about 2 with respect to each other model.

\section{Conclusion and discussion}

This paper is mainly on methodological issues related 
to model comparisons in critical,
frontier physics cases where the prior knowledge is relevant.

We have given reasons of why `conventional statistics'
(i.e. the collection of frequentistic prescriptions)
does not adequately approach the problem of model comparison,
mainly because of impossibility of classify hypotheses in 
a probability scale and of the pretension that good criteria
to state what is `significant' can be derived 
from the properties of the null hypothesis alone, 
without considering the details of the alternative 
hypotheses.\footnote{If you are puzzled by the
question ``why do frequentistic
hypothesis test often work?'', you might give a look 
at Chap. 10 of Ref.~\cite{giulioWS}.}

Within the so called  Bayesian framework, we have 
started from the basic observation 
that the most a  probability theory 
should do is to provide
rules to modify our beliefs on the light of experimental data.
Beliefs can be about the values of the parameters of a model or 
about alternative models. As far as beliefs on model parameters
are concerned, we have shown that the likelihood, rescaled to its
insensitivity limit value (the ${\cal R}$ function, 
or `relative belief 
updating factor'), represents 
a good, prior independent way of summarizing the information
contained in the data with respect to a given model.
Indeed, when this method is applied to the Explorer-Nautilus data,
from the visual inspection of the ${\cal R}$ function
the reader gets, for each model, an immediate 
overview of what the data say about 
the number of events involved in the observation.
The $\alpha$ values for which the relative belief 
updating factor is maximum correspond to a total number of
g.w. events in the data ($n_{gwc}$) about 10 for all three Galactic 
models. For those who share beliefs 
that numbers of this order
of magnitude or more are possible
(and that one of the three models is the correct one), 
the ${\cal R}$ can be translated 
into a result $n_{gwc}\approx 10\pm 5$ (or a rate on Earth of $\approx 1.0\pm0.5$).

Going to the model comparison, we have shown the unavoidable complication
due to the fact that each model depends on a free parameter ($\alpha$) 
and, hence, the Bayes factors depend on the prior pdf of this parameter,
i.e. $f_{\circ_{\cal M}}(\alpha)$. 
Since the models used do not come with a kind of reference $f_{\circ_{\cal M}}(\alpha)$
(we hope that more work will be done in this direction) 
we had to do some choices and we have given the results under
different scenarios, from the most negative one (``there is no chance that
the models produce something observable, given the present energy
sensitivity'') to some others in which $\alpha$ above 1
are conceivable (described by the priors 
we have called in the text `moderate' and 'uniform'). Given these
scenarios the Galactic Center model gets preferred
over the others by a Bayes factor of about 2:1. 

We would like to end replying to the
objection, arisen often in discussions, that 
``the plot with coincidences grouped in bins of sidereal time
provides the same information of that in which 
coincidences are grouped in bins of solar time''.
This might be true if one is blindly looking for ``statistical significance'',
following strictly frequentistic prescriptions, which, as explained above,
we do not consider the proper way to go. 
%%[many igNobel prices should have been given in the past decades
%%to (especially particle-)physicists 
%%who finally found a good ``statistical significance''\ldots]. 

To answer this objection we have done the exercise of applying
exactly the same analysis with the same models to the data
grouped in bins of solar time.
 \begin{figure}
\begin{center}
\epsfig{file=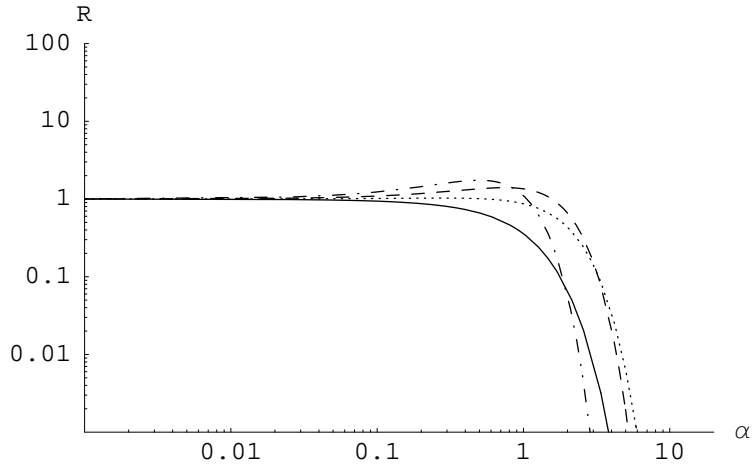,clip=,width=10.cm}
\caption{Same as Fig.~\ref{fig:Rall05}, but for data grouped in solar time bins}.
\label{fig:Rsolare}
\end{center}
\end{figure}
\begin{table}
\begin{center}
\caption{Same as Tab.~\ref{tab:due05}, but for data grouped in solar time bins}
\label{tab:tre}
\begin{tabular}{cccc} 
\hline 
 &  `sceptical'  &  `moderate'  &  `uniform' \\
$
\begin{array}{lr}  
& f_0(\alpha) \\
 & \\ 
 \mbox{Model} & 
\end{array}
$
 & \epsfig{file=figlavoropub/figskeptical.eps,clip=,width=2.5cm,height=1.0cm} 
&\epsfig{file=figlavoropub/figmoderate.eps,clip=,width=2.5cm,height=1.0cm}
&\epsfig{file=figlavoropub/figoptimistic.eps,clip=,width=2.5cm,height=1.0cm}\\
\hline
%\multicolumn{1}{c}{GC} & & & \\
\epsfig{file=figlavoropub/figsignalCG.eps,clip=,width=2.5cm} 
  &  {\huge 1.0}    &{\huge  0.5}    &{\huge  0.1}   \\
\hline
%\multicolumn{1}{c}{GD} & & & \\
\epsfig{file=figlavoropub/figsignalDG.eps,clip=,width=2.5cm}   & {\huge 1.1}   & {\huge 1.2} &{\huge 0.3}   \\ 
\hline 
%\multicolumn{1}{c}{GMD} & & & \\
\epsfig{file=figlavoropub/figsignalM1.eps,clip=,width=2.5cm} &   {\huge 1.0}     &{\huge  0.9}  & {\huge 0.2}   \\ 
\hline
%\multicolumn{1}{c}{UNI} & & & \\
\epsfig{file=figlavoropub/figsignalUNIF.eps,clip=,width=2.5cm} &   {\huge 1.2}     &{\huge  1.2}  & {\huge 0.2}   \\ 
\hline 
\end{tabular}
\end{center}
\end{table}
The results are given in
Fig.~\ref{fig:Rsolare} and Tab.~\ref{tab:tre}, which have the
analogous meaning of Fig.~\ref{fig:Rall05} and Tab.~\ref{tab:due05}. 
The ${\cal R}$ functions do not show values of $\alpha$ particularly
preferred by the data: they have approximately a smoothed 
step-function shape which divides the orders of magnitudes of $\alpha$
(on the right side) that are excluded from those in which the experiment 
loses sensitivity. This is more ore less what we could get distributing
34 entries in 48 bins according to a multinomial distribution:
small differences in the shape of ${\cal R}_{\cal M}(\alpha)$
depend on the individual occurrence of the multinomial data set
(but never forget that the multinomial distribution does not
{\it forbid} strong clustering of the entries around one bin!).

The lesson from this exercise is that the Explorer-Nautilus 2001 data,
plotted as a function of a sensible physical quantity and compared with 
physically motivated models, does not provide the same information
of any random sample. Indeed, the evidence in support of the
models is not enough to modify strongly our beliefs, 
but it is certainly at the level of ``stay tuned'', waiting 
for results of the 2003 run.

\section{Acknowledgments}
The authors thank the ROG collaboration for having provided the 2001 data
of Explorer and Nautilus.
We also thank  G. Giordano who has given us the
information to compute the mass distribution model of Galaxy (GMD).
Finally, P.A. and S.D. thank warmly the organizers of GWDAW for such
a vital and fruitful workshop.

\end{document}